 \documentclass[final,review,3p,times,authoryear]{elsarticle}
\usepackage{amsmath}
\usepackage{amssymb}
\usepackage{color}
\usepackage{array}
\usepackage{enumerate}
\usepackage{multirow}
\usepackage{lipsum}
\usepackage{algorithm2e}

\journal{Computational Statistics and Data Analysis}

\begin{document}

\begin{frontmatter}



\title{A penalized simulated maximum likelihood approach in parameter estimation for stochastic differential equations}


\author[csu]{Libo Sun}\ead{sun@stat.colostate.edu}
\author[csu]{Chihoon Lee\corref{cor1}}\ead{chihoon@stat.colostate.edu}
\author[csu]{Jennifer A. Hoeting} \ead{jah@stat.colostate.edu}

\address[csu]{Department of Statistics, Colorado State University, Fort Collins, Colorado 80523, U.S.A.}

\cortext[cor1]{Corresponding author. Tel: +1(970) 491-7321}

\begin{abstract}
We consider the problem of estimating parameters of stochastic differential equations (SDEs) with discrete-time observations that are either completely or partially observed. The transition density between two observations is generally unknown. We propose an importance sampling approach with an auxiliary parameter when the transition density is unknown. We embed the auxiliary importance sampler in a penalized maximum likelihood framework which produces more accurate and computationally efficient parameter estimates. Simulation studies in three different models illustrate promising improvements of the new penalized simulated maximum likelihood method. The new procedure is designed for the challenging case when some state variables are unobserved and moreover, observed states are sparse over time, which commonly arises in ecological studies. We apply this new approach to two epidemics of chronic wasting disease in mule deer.
\end{abstract}




\end{frontmatter}



\section{Introduction}

It is very important for ecologists and wildlife managers to understand the dynamics of infectious diseases, such as chronic wasting disease (CWD) which is a fatal contagious disease in cervid populations \citep{miller2006dynamics}. Several ordinary differential equation models have been proposed by \citet{miller2006dynamics} to describe the transmission mechanism of CWD. Stochastic epidemic models allow more realistic description of the transmission of disease as compared to deterministic epidemic models \citep{becker1979uses, andersson2000stochastic}. However, parameter estimation is challenging for discretely observed data for stochastic models \citep{sorensen2004parametric, jimenez2005inference}. Stochastic differential equation (SDE) models are a natural extension of ordinary differential equation models and they may be simpler to derive and apply than Markov chain models. For example, the transition matrix in Markov chain models can be very complicated when there are several interacting populations \citep{allen2003comparison, allen2005comparison}. Moreover, SDEs have broader application areas, which include not only ecology and biology but also economics, finance, bioinformatics, and engineering. 

Various methods for inferential problems for SDEs have been developed. The Hermite polynomial expansion approach proposed by \citet{ait2002transition, ait2008closed} may perform poorly if the data are sparsely sampled \citep{stramer2007simulated}. Moreover, this approach has some restrictions which could limit its application, especially for multivariate models \citep{lindstrom2012regularized}. \citet{sarkka2008application} proposed an approach which uses an alternative SDE as an importance process and the Girsanov theorem to help evaluate the likelihood ratios of two SDEs. However, the diffusion coefficient of their model is state-independent, whereas general SDE models allow for a state-dependent diffusion coefficient. Recent developments have mainly been focused on Bayesian approaches \citep{eraker2001mcmc, golightly2005bayesian, golightly2006bayesian, golightly2011bayesian, donnet2010bayesian}, which can suffer a very slow rate of convergence as the dimension of the model increases and the data are sparsely sampled. We propose a penalized simulated maximum likelihood (PSML) approach which is computationally feasible.

For a SDE model the transition density between two observations is known in only a few univariate cases. \citet{pedersen1995new} firstly proposed a simulated maximum likelihood (SML) approach which integrates out the unobserved states using Monte Carlo integration with importance sampling. We refer to the basic sampler in this approach as the Pedersen sampler. Although the Pedersen sampler may provide estimates that are arbitrarily close to the true transition density, it is computationally expensive in practice. \citet{durham2002numerical} proposed several different importance samplers in a SML framework to improve the efficiency of the Pedersen sampler. They concluded their modified Brownian bridge (MBB) sampler has the best performance in terms of accuracy in root mean square error and efficiency in time. \citet{richard2007efficient} proposed an efficient importance sampling technique which converts the problem of minimizing the variance of an approximate likelihood to a recursive sequence of auxiliary least squares optimization problems. \citet{pastorello2010efficient} applied Richard and Zhang's approach to estimate the parameters of some univariate SDE models. However, the extension to multivariate SDEs with partially observed data is not trivial. \citet{lindstrom2012regularized} introduced a regularized bridge sampler, which is a weighted combination of the Pedersen sampler and the MBB sampler, for sparsely sampled data.

The methods of \citet{pedersen1995new} and \citet{durham2002numerical} have mainly been applied in the area of econometrics. Here we propose a methodology to improve the MBB sampler and the regularized sampler and extend them to the area of ecology. From an inferential viewpoint, practitioners must contend with two major challenges: (a) in the multivariate state space, some state variables are completely unobserved; (b) observed data are quite sparse over time. These are common features of ecological data. For example, the number of deaths for CWD in a wild animal population can be observed or estimated, but the numbers of infected and susceptible animals may be impossible or costly to obtain. Moreover, the time interval between two consecutive observations could be very long, usually weeks or even months. With such partially observed sparse data, the MBB approach no longer has the same promising results as in the univariate case. Although the regularized sampler in \citet{lindstrom2012regularized} is designed for sparsely sampled data, the optimal choice of the weight parameter $\rho$ (which is denoted as $\alpha$ in the cited paper) needs to be determined. We propose an importance sampling approach with an auxiliary parameter which provides more accurate estimates of the parameters of an SDE when the transition density is unknown.We embed the auxiliary importance sampler in a penalized maximum likelihood framework. The penalty term we add to the log likelihood is a constraint on selecting the importance sampler. We show via simulation studies that our approach improves the accuracy of parameter estimates and computational efficiency compared to the MBB sampler and the regularized sampler.

The remainder of the paper is organized as follows. In Section 2, we present the general multivariate SDE model. Section 3 provides brief descriptions of the Pedersen, MBB  and regularized samplers. Section 4 describes our methodology in detail. Section 5 presents simulation studies for different models. Section 6 illustrates our method on a CWD dataset as a real world example.  Section 7 concludes with a discussion.

\section{Background}
We begin with the general multivariate SDE model where some state variables are unobserved. Let $\boldsymbol{X}(t)=\{X_1(t),\ldots,X_k(t)\}^T$ denote a $k$-dimensional state variable vector at time $t\geq 0$. Consider a multivariate SDE model, 
\begin{equation}\label{general_SDE}
d\boldsymbol{X}(t)=f(\boldsymbol{X}(t),\boldsymbol{\theta})dt+g(\boldsymbol{X}(t),\boldsymbol{\theta})d\boldsymbol{W}(t)
\end{equation} with known initial condition $\boldsymbol{X}(t_0)=\boldsymbol{x}_0$,
where $\boldsymbol{\theta}\in\Theta\subseteq\mathbb{R}^p$ is an unknown $p$-dimensional parameter vector, $\boldsymbol{W}$ is a $k$-dimensional standard Wiener process, and both functions $f: \mathbb{R}^k\times\Theta\rightarrow\mathbb{R}^k$ and $g: \mathbb{R}^k\times\Theta\rightarrow\mathbb{R}^{k\times k}$ are known. Note that the derivation below still holds for the case with unknown initial condition $\boldsymbol{X}(t_0)$, which can be treated as another unknown parameter. We assume that the SDE \eqref{general_SDE} has a unique weak solution. See \citet[Chapter~5]{oksendal2010stochastic} for conditions that ensure this. 

We assume that only a subset of the state process $\{\boldsymbol{X}_{\text{obs}}(t)\}_{t\geq0}$ can be observed at discrete time points. It is natural to suppose only $\boldsymbol{X}_{\text{obs}}(t_i)=\{X_j(t_i),\ldots,X_k(t_i)\}$ is observed at $t_i$, for $1< j\leq k$ and $i=1,\ldots,n$, and all other state variables $\boldsymbol{X}_{-\text{obs}}(t_i)=\{X_1(t_i),\ldots, X_{j-1}(t_i)\}$, are unobserved. In the case of complete observation, that is when $j=1$, a similar derivation as below can be obtained. Note that time intervals do not have to be equidistant.

The discrete-time likelihood of model 
\eqref{general_SDE} is given by 
\begin{align}\label{likelihood_sim}
L(\boldsymbol{\theta}) = p(\boldsymbol{X}_{\text{obs}}(t_1)|\boldsymbol{X}(t_0),\boldsymbol{\theta})\prod_{i=2}^n p(\boldsymbol{X}_{\text{obs}}(t_i)|\boldsymbol{X}(t_0),\boldsymbol{X}_{\text{obs}}(t_1:t_{i-1}),\boldsymbol{\theta})
\end{align}
where $\boldsymbol{X}_{\text{obs}}(t_1:t_{i-1})$ denotes all observations of $\boldsymbol{X}_{\text{obs}}$ from time $t_1$ to $t_{i-1}$. We omit the parameter $\boldsymbol{\theta}$ for brevity from now on. Notice that the term $p(\boldsymbol{X}_{\text{obs}}(t_i)|\allowbreak \boldsymbol{X}(t_0),\allowbreak \boldsymbol{X}_{\text{obs}}(t_1:t_{i-1}))$ is not available in closed form except for simple cases. However, factoring the likelihood as in \eqref{likelihood_sim} allows us to evaluate the likelihood given by
\begin{align*}
&p(\boldsymbol{X}_{\text{obs}}(t_i)|\boldsymbol{X}(t_0),\boldsymbol{X}_{\text{obs}}(t_1:t_{i-1}))=\int p(\boldsymbol{X}_{\text{obs}}(t_i)|\boldsymbol{X}(t_{i-1}))p(\boldsymbol{X}_{-\text{obs}}(t_{i-1})|\boldsymbol{X}(t_0),\boldsymbol{X}_{\text{obs}}(t_1:t_{i-1})) d\boldsymbol{X}_{-\text{obs}}(t_{i-1}). 
\end{align*}
A feasible approach to evaluate this integral is via Monte Carlo integration. That requires a method to draw samples from the distribution of $\boldsymbol{X}_{-\text{obs}}(t_{i-1})|\boldsymbol{X}(t_0),\boldsymbol{X}_{\text{obs}}\allowbreak(t_1:t_{i-1})$. It can be shown that (cf. \citeauthor{durham2002numerical}, 2002)
\begin{align}
&p(\boldsymbol{X}_{-\text{obs}}(t_i)|\boldsymbol{X}(t_0),\boldsymbol{X}_{\text{obs}}(t_1:t_i))\propto\int p(\boldsymbol{X}(t_i)|\boldsymbol{X}(t_{i-1}))p(\boldsymbol{X}_{-\text{obs}}(t_{i-1})|\boldsymbol{X}(t_0),\boldsymbol{X}_{\text{obs}}(t_1:t_{i-1}))d\boldsymbol{X}_{-\text{obs}}(t_{i-1}), \label{tran_3}
\end{align}
for $i=1,...,n$. Therefore, assuming $p(\boldsymbol{X}(t_i)|\boldsymbol{X}(t_{i-1}))$ is known (see below), iterative application of Monte Carlo integration \eqref{tran_3} yields an approximation of $\boldsymbol{X}_{-\text{obs}}(t_\ell)|\allowbreak\boldsymbol{X}(t_0),\boldsymbol{X}_{\text{obs}}(t_1:t_\ell)$ for $\ell\geq 1$. This is similar in spirit to a particle filter \citep{durham2002numerical, pitt1999filtering}, but our model does not include measurement errors. The algorithmic form of this simple sequential Monte Carlo algorithm is provided in \ref{append1}.

It is left to approximate the transition probability density $p(\boldsymbol{X}(t_i)|\boldsymbol{X}(t_{i-1}))$ which has no closed form in most cases. The Euler-Maruyama scheme \citep{kloeden1992numerical} is a common approach to approximate the solution of an SDE, which is given by
\begin{align}\label{general_SDE_euler}
\boldsymbol{X}&(t+\delta)-\boldsymbol{X}(t)\approx f(\boldsymbol{X}(t),\boldsymbol{\theta})\delta+g(\boldsymbol{X}(t),\boldsymbol{\theta})(\boldsymbol{W}(t+\delta)-\boldsymbol{W}(t)), 
\end{align}
where $\delta$ is the step size and $\bold{W}(t+\delta)-\bold{W}(t)$ follows a multivariate normal distribution with variance matrix $\delta\boldsymbol{\mathcal{I}}_{k\times k}$, where $\boldsymbol{\mathcal{I}}$ is the identity matrix. This Euler-Maruyama scheme works well if the step size is small. Hence, if the time interval between two observations is small enough, we can approximate $p(\boldsymbol{X}(t_i)|\boldsymbol{X}(t_{i-1}))$ using a multivariate normal density.

If the time interval between observations is large, the above approximation will introduce bias. We can partition the interval $t_{i-1}$ to $t_i$ to $M$ subintervals such that $\delta=(t_i-t_{i-1})/M$ is small enough for the Euler-Maruyama scheme. By the Markov property,  \citet{pedersen1995new} proved that $p(\boldsymbol{X}(t_i)|\boldsymbol{X}(t_{i-1}))$ can be approximated by
\begin{align}
\label{importance_sampler}
&p^{(M)}\big(\boldsymbol{X}(t_i)|\boldsymbol{X}(t_{i-1})\big)\equiv\int\prod_{m=1}^{M}p^{(1)}\big(\boldsymbol{X}(t_{i-1}+m\delta)|\boldsymbol{X}(t_{i-1}+(m-1)\delta)\big)d\boldsymbol{X}\big((t_{i-1}+\delta):(t_i-\delta)\big),
\end{align}
where $p^{(1)}(\cdot)$ is the multivariate normal density approximated by Euler-Maruyama scheme.

Then, our goal is to compute $p^{(M)}(\boldsymbol{X}(t_i)|\boldsymbol{X}(t_{i-1}))$. Using importance sampling, we draw i.i.d. $J$ samples, $\{\boldsymbol{X}^{(j)}((t_{i-1}+\delta):(t_i-\delta)), j=1,\cdots,J\}$, from an importance sampler $q$, then \eqref{importance_sampler} can be approximated by
\begin{equation}
\label{importance_sampler_MC}
\frac{1}{J}\sum_{j=1}^Jh\Big(\boldsymbol{X}^{(j)}\big((t_{i-1}+\delta):(t_i-\delta)\big)\Big),
\end{equation}
where 
\begin{align}\label{IS_h}
h\Big(\boldsymbol{X}^{(j)}\big((t_{i-1}+&\delta):(t_i-\delta)\big)\Big)\equiv\frac{\prod_{m=1}^{M}p^{(1)}\big(\boldsymbol{X}^{(j)}(t_{i-1}+m\delta)|\boldsymbol{X}^{(j)}(t_{i-1}+(m-1)\delta)\big)}{q\big(\boldsymbol{X}^{(j)}((t_{i-1}+\delta):(t_i-\delta))\big)}.
\end{align}
The convergence of the importance sampling estimator \eqref{importance_sampler_MC} to \eqref{importance_sampler} as $J
\rightarrow\infty$ is shown by \citet{geweke1989bayesian}. The estimator \eqref{importance_sampler_MC} is an unbiased estimator, regardless of the choice of the importance sampler $q$. The variance of \eqref{importance_sampler_MC} is given by
\begin{align}\label{importance_sampler_variance}
&\text{Var}\left(\frac{1}{J}\sum_{j=1}^Jh\left(\boldsymbol{X}^{(j)}\big((t_{i-1}+\delta):(t_i-\delta)\big)\right)\right)=\frac{1}{J}\text{Var}\bigg(h\Big(\boldsymbol{X}\big((t_{i-1}+\delta):(t_i-\delta)\big)\Big)\bigg)\nonumber\\
&=\frac{1}{J}\left[\int\frac{\prod_{m=1}^{M}\Big[p^{(1)}\big(\boldsymbol{X}(t_{i-1}+m\delta)|\boldsymbol{X}(t_{i-1}+(m-1)\delta)\big)\Big]^2}{q\big(\boldsymbol{X}((t_{i-1}+\delta):(t_i-\delta))\big)}d\boldsymbol{X}\big((t_{i-1}+\delta):(t_i-\delta)\big)-\big[p^{(M)}(\boldsymbol{X}(t_i)|\boldsymbol{X}(t_{i-1}))\big]^2\right],
\end{align}
which attains its minimum of $0$ when 
\begin{align}\label{best_q}
q\big(\boldsymbol{X}((t_{i-1}+\delta)&:(t_i-\delta))\big)=\frac{\prod_{m=1}^{M}p^{(1)}(\boldsymbol{X}(t_{i-1}+m\delta)|\boldsymbol{X}(t_{i-1}+(m-1)\delta))}{p^{(M)}(\boldsymbol{X}(t_i)|\boldsymbol{X}(t_{i-1}))}.
\end{align}
Thus in theory a single sample is sufficient to approximate $p^{(M)}(\boldsymbol{X}(t_i)|\boldsymbol{X}(t_{i-1}))$. However, in practice this is infeasible because $p^{(M)}(\boldsymbol{X}(t_i)|\boldsymbol{X}(t_{i-1}))$ is unknown. 

In order to decrease the variance \eqref{importance_sampler_variance} and reduce the sample size $J$, we want to choose an importance sampler $q\big(\boldsymbol{X}((t_{i-1}+\delta):(t_i-\delta))\big)$ that is as close as possible to $\prod_{m=1}^{M}p(\boldsymbol{X}(t_{i-1}+m\delta)|\boldsymbol{X}(t_{i-1}+(m-1)\delta))$, which is the principle of choosing the proposal density in importance sampling. 

\section{Importance samplers for simulated maximum likelihood}
Here we review three importance samplers for approximating the transition probability density $p^{(M)}(\boldsymbol{X}(t_i)|\allowbreak\boldsymbol{X}(t_{i-1}))$ in \eqref{importance_sampler}. These approaches can be used to compute maximum likelihood estimates of the parameters of the SDE model \eqref{general_SDE} (i.e., simulated maximum likelihood estimation). In Section \ref{PSML} we propose a new penalized simulated maximum likelihood approach which can be used to improve the performance of all three methods described below.

\subsection{Pedersen sampler}
The Pedersen sampler is the first importance sampler proposed to approximate a transition density \citep{pedersen1995new, santa1997simulated}. The Pedersen sampler constructs the importance sampler $q$ by simulating $J$ paths on each subinterval just using the Euler-Maruyama scheme conditional on $\boldsymbol{X}(t_{i-1})$, so the first $M-1$ terms in \eqref{IS_h} are canceled. Hence, \eqref{importance_sampler_MC} reduces to
\begin{equation}\label{Pedersen_sampler}
\frac{1}{J}\sum_{j=1}^J p^{(1)}(\boldsymbol{X}(t_i)|\boldsymbol{X}^{(j)}(t_i-\delta)).
\end{equation}

One can simulate $J$ trajectories of all $k$-dimensional state process $\boldsymbol{X}$ from time $t_{i-1}$ to time $t_i-\delta$ by using the Euler-Maruyama scheme with the step size $\delta$. Although the Pedersen sampler has a very simple form, it is well known that it is computationally intensive in practice \citep{durham2002numerical}, especially for a multivariate SDE model. The Pedersen sampler can introduce excessive variance in the simulation of all possible transition probabilities even with a very large number of simulated trajectories. 

\subsection{Modified Brownian bridge sampler}
A more efficient importance sampler is called the modified Brownian bridge (MBB) sampler, which is originally proposed by \citet{durham2002numerical} for the univariate case and modified by \citet{golightly2006bayesian} for the multivariate case. Instead of simulating a path on each subinterval using the Euler approximation  based on $\boldsymbol{X}(t_{i-1})$ as in Pedersen sampler, this method draws $\boldsymbol{X}((t_{i-1}+\delta):(t_i-\delta))$ conditional on $\boldsymbol{X}(t_{i-1})$ and $\boldsymbol{X}_{\text{obs}}(t_i)$. Here, we outline the procedure. See \citet{golightly2006bayesian} for more details.

Let $\boldsymbol{X}^m$ denote $\boldsymbol{X}(t_{i-1}+m\delta)$ and partition the drift and diffusion functions in \eqref{general_SDE} as
\begin{equation*}
f(\boldsymbol{X})=\begin{pmatrix}
f_{-\text{obs}}(\boldsymbol{X})\\
f_{\text{obs}}(\boldsymbol{X})
\end{pmatrix}
\end{equation*}
and 
\begin{equation*}
g^T(\boldsymbol{X})g(\boldsymbol{X})=\begin{bmatrix}
G_{-\text{obs},-\text{obs}}(\boldsymbol{X}) & G_{-\text{obs},\text{obs}}(\boldsymbol{X})\\
G_{\text{obs},-\text{obs}}(\boldsymbol{X}) & G_{\text{obs},\text{obs}}(\boldsymbol{X})\\
\end{bmatrix}.
\end{equation*}
Then the MBB sampler draws $\boldsymbol{X}^{m+1}$ from the density
\begin{equation}\label{bridge_density}
q(\boldsymbol{X}^{m+1}|\boldsymbol{X}^m,\boldsymbol{X}_{\text{obs}}(t_i))=\phi(\boldsymbol{X}^{m+1};\boldsymbol{X}^m+\boldsymbol{\eta}_m\delta, \Sigma_m\delta),
\end{equation} 
where $\phi(\boldsymbol{X}; \boldsymbol{\mu},\Sigma)$ is a multivariate normal density with mean vector $\boldsymbol{\mu}$ and covariance matrix $\Sigma$. Here
\begin{equation}\label{MBB_eta}
\boldsymbol{\eta}_m=\begin{pmatrix}
f_{-\text{obs}}(\boldsymbol{X}^m)+\frac{G_{-\text{obs},\text{obs}}(\boldsymbol{X}^m)}{\delta(M-m)G_{\text{obs},\text{obs}}(\boldsymbol{X}^m)}\Delta_\text{\text{obs}}\\
(\boldsymbol{X}_{\text{obs}}(t_i)-\boldsymbol{X}_{\text{obs}}(t_{i-1}+m\delta))/[\delta(M-m)]\\
\end{pmatrix},
\end{equation} 
and 
\begin{align}\label{MBB_sigma}
\Sigma_m=\begin{bmatrix}
G_{-\text{obs},-\text{obs}}(\boldsymbol{X}^m)-\frac{G_{-\text{obs},\text{obs}}(\boldsymbol{X}^m)G_{\text{obs},-\text{obs}}(\boldsymbol{X}^m)}{(M-m)G_{\text{obs},\text{obs}}(\boldsymbol{X}^m)} &\frac{M-m-1}{M-m}G_{-\text{obs},\text{obs}}(\boldsymbol{X}^m)\\
\frac{M-m-1}{M-m}G_{\text{obs},-\text{obs}}(\boldsymbol{X}^m) & \frac{M-m-1}{M-m}G_{\text{obs},\text{obs}}(\boldsymbol{X}^m)
\end{bmatrix},
\end{align} 
where 
\begin{equation*}
\Delta_{\text{\text{obs}}}=\boldsymbol{X}_{\text{obs}}(t_i)-\left[\boldsymbol{X}_{\text{obs}}(t_{i-1}+m\delta)+f_{\text{obs}}(\boldsymbol{X}^m)(M-m-1)\delta\right]
\end{equation*}
for $m=0,1,\cdots,M-2$. For $m=M-1$, we draw $\boldsymbol{X}_{-\text{obs}}(t_i)$ conditional on $\boldsymbol{X}^{M-1}=\boldsymbol{X}(t_i-\delta)$ and $\boldsymbol{X}_{\text{obs}}(t_i)$, which is conditional multivariate normal by the Euler-Maruyama scheme. By recursively drawing from \eqref{bridge_density} one can obtain a Brownian bridge, $\boldsymbol{X}((t_{i-1}+\delta):(t_i-\delta))$ conditioned on starting at $\boldsymbol{X}(t_{i-1})$ and finishing at $\boldsymbol{X}_{\text{obs}}(t_i)$.

\subsection{Regularized sampler}
The MBB sampler can produce a poor approximation because its linear interpolation between two observations ignores the dynamics of the model in constructing the sample paths, especially when the diffusion dynamics are dominated by the drift term for sparsely sampled data \citep{lindstrom2012regularized}. A regularized sampler which is a weighted combination of the Pedersen sampler and the MBB sampler is proposed by \citet{lindstrom2012regularized} to overcome this limitation. Here we give the explicit form of this regularized sampler.

Let $\boldsymbol{\mu}_P$ and $\Sigma_P$ be the mean and the variance of the Pedersen sampler and $\boldsymbol{\mu}_M$ and $\Sigma_M$ be the mean and the variance of the MBB sampler. Then the regularized sampler draws $\boldsymbol{X}^{m+1}$ from the density
\begin{align}\label{regularized_sampler}
q_\rho(\boldsymbol{X}^{m+1}&|\boldsymbol{X}^m,\boldsymbol{X}_{\text{obs}}(t_i))=\phi(\boldsymbol{X}^{m+1};\left(\mathcal{I}-\boldsymbol{V}\right)\boldsymbol{\mu}_P+\boldsymbol{V}\boldsymbol{\mu}_M, \left(\mathcal{I}-\boldsymbol{V}\right)\boldsymbol{\Sigma}_P+\boldsymbol{V}\boldsymbol{\Sigma}_M),
\end{align} 
where $\mathcal{I}$ is the identity matrix and 
\begin{equation}\label{matrix_V}
\boldsymbol{V}=\frac{M-m}{M-m+\rho(M-m-1)^2}\mathcal{I},
\end{equation}
where $\rho\in[0,1]$. The regularized sampler is dominated by the Pedersen sampler initially and is dominated by the MBB sampler as $m\rightarrow (M-1)$ in \eqref{matrix_V}. The regularized sampler depends on the parameter $\rho$. A large $\rho$ will make the regularized sampler similar to the Pedersen sampler and a smaller $\rho$ will make it similar to the MBB sampler. \citet{lindstrom2012regularized} used $\rho=0.1$ throughout, however, did not propose an algorithm for choosing the optimal $\rho$. Hence, a practical algorithm for selecting the optimal $\rho$ is needed for successful implementation of the regularized sampler. We propose one such approach in the next section.

\section{Penalized simulated maximum likelihood and auxiliary importance sampling}\label{PSML}
To find an efficient importance sampler, we need to minimize \eqref{importance_sampler_variance}, the variance of the approximation of the transition density. Here we propose a new approach to minimize the variance; (i) we augment the likelihood with an auxiliary parameter $\rho$ which tunes the importance sampler to the model parameters and (ii) we maximize the log likelihood with a constraint on the coefficient of variation of the importance sampler.

\subsection{Penalized simulated maximum likelihood}
In our penalized simulated maximum likelihood (PSML) approach, we maximize the log likelihood subject to the sum of the coefficient of variation of the Monte Carlo approximation of the transition density being less than a prespecified level. Suppose a family of auxiliary importance samplers $\{q_\rho\}$ has been selected, where $\rho$ is an auxiliary or nuisance parameter. Our goal is to find $\hat{\rho}$ that minimizes the sum of the coefficient of variation of the Monte Carlo approximation of the transition density. 

Let $h_\rho$ be the importance sampling weights to approximate $p(\boldsymbol{X}_{\text{obs}}(t_i)|\boldsymbol{X}(t_0),\boldsymbol{X}_{\text{obs}}(t_1:t_{i-1}))$ in \eqref{general_SDE}. Specifically,
\begin{align}
\label{h_rho}
h_\rho\Big(\boldsymbol{X}^{(j)}&(t_{i-1}:(t_i-\delta))\Big)\equiv\frac{p^{(1)}(\boldsymbol{X}_{\text{obs}}(t_i)|\boldsymbol{X}^{(j)}(t_i-\delta))\prod_{m=1}^{M-1}p^{(1)}(\boldsymbol{X}^{(j)}(t_{i-1}+m\delta)|\boldsymbol{X}^{(j)}(t_{i-1}+(m-1)\delta))}{q_\rho(\boldsymbol{X}^{(j)}(t_{i-1}:(t_i-\delta)))},
\end{align}
where $\boldsymbol{X}^{(j)}(t_{i-1})\equiv\{\boldsymbol{X}_{-\text{obs}}^{(j)}(t_{i-1}),\boldsymbol{X}_{\text{obs}}(t_{i-1})\}$ and $q_\rho$ is the importance sampler density, e.g., \eqref{regularized_sampler} and \eqref{auxiliary_importance} below. We adopt the notation $h_\rho$ to indicate the expression in \eqref{h_rho}, suppressing the dependence on $i$ and $j$ for notational simplicity. The PSML estimator $(\hat{\boldsymbol{\theta}},\hat{\rho})$ is defined by 
\begin{align}\label{pen_max}
(\hat{\boldsymbol{\theta}},\hat{\rho})=&\text{arg max}\sum_{i=1}^n \log\left(\frac{1}{J}\sum_{j=1}^Jh_\rho\right)\text{ subject to } \sum_{i=1}^n \widehat{\text{cv}}\left(h_\rho\right)\leq s, 
\end{align}
where $s\geq 0$ is a tuning parameter and $\widehat{\text{cv}}(h_\rho)$ is the sample coefficient of variation of $h_\rho$, which is the sample standard deviation of the $J$ importance weights $h_\rho$ divided by their sample mean. Notice that \eqref{pen_max} is reminiscent of LASSO \citep{tibshirani1996regression}, and is equivalent to maximizing a penalized log likelihood,
\begin{equation}\label{pen_max_lambda}
l^*(\boldsymbol{\theta},\rho)=\sum_{i=1}^n \log\left(\frac{1}{J}\sum_{j=1}^Jh_\rho\right) -\lambda\sum_{i=1}^n \widehat{\text{cv}}(h_\rho), 
\end{equation}
where $\lambda$ in \eqref{pen_max_lambda} has a one-to-one mapping to $s$ in \eqref{pen_max}. The reason the coefficient of variation is chosen instead of the variance is that the former is a normalized measurement which is not affected by the magnitude of the data. This makes it easier to choose the tuning parameter $\lambda$ in practice, as will be shown below. When the penalty term is omitted, the parameter estimates have a large variance because the importance sampler is not well tuned.

The constraint, $\sum_{i=1}^n\widehat{\text{cv}}(h_\rho)\leq s$ in \eqref{pen_max}, is equivalent to a constraint on the effective sample size \citep[Chapter~6]{givens2012computational}, $$\widehat{N}(q_\rho,p)\equiv\frac{J}{1+\frac{1}{n}\sum_{i=1}^n\hat{\text{cv}}^2(h_\rho)}\geq\frac{J}{1+\frac{s^2}{n}}.$$ The effective sample size measures how much the auxiliary importance sampler density $q_\rho$ differs from the target density $p$, and it can be interpreted as $J$ weighted samples are worth $\widehat{N}(q_\rho,p)$ unweighted i.i.d. samples drawn exactly from target density $p$. Effective sample size can be used as a measure of computational efficiency.

The tuning parameter $s$ controls how close the auxiliary importance sampler density $q_\rho$ is to the product of transition probability densities, the numerator of \eqref{IS_h}. Let $s^0$ denote the sum of the coefficient of variation for the approximation of the transition density by the Pedersen sampler. When $s<s^0$ the resulting auxiliary importance sampler will have smaller variance \eqref{importance_sampler_variance} than that from the Pedersen sampler \eqref{Pedersen_sampler}. When $s=0$, the constraint in \eqref{pen_max} requires that the auxiliary importance sampler $q_\rho$ attains its ideal case \eqref{best_q}. However, as $s\rightarrow 0$, $\lambda\rightarrow\infty$ and therefore the log likelihood plays no role in estimating $\boldsymbol{\theta}$.

The tuning parameter $\lambda$ in \eqref{pen_max_lambda} can be estimated using various techniques. We choose the value that minimizes the estimated prediction error,
\begin{equation}\label{Estimated_PE}
\epsilon_\lambda\equiv\frac{1}{nL}\sum_{\ell=1}^L\sum_{i=1}^n\lVert\widehat{\boldsymbol{X}}^{(\ell)}_{\text{obs}}(t_i)-\boldsymbol{X}_{\text{obs}}(t_i)\rVert,
\end{equation}
where $\widehat{\boldsymbol{X}}^{(\ell)}_{\text{obs}}(t_i)$ is the $\ell$th simulated $\boldsymbol{X}_{\text{obs}}$ at observation time $t_i$ by the Euler-Maruyama scheme \eqref{general_SDE_euler} with $\boldsymbol{\theta}=\widehat{\boldsymbol{\theta}}(\lambda)$ and $\lVert\boldsymbol{X}\rVert$ is a Euclidean norm of $\boldsymbol{X}$ in $\mathbb{R}^k$. The number of simulations $L$ is chosen arbitrarily and is set to 1000 here. More details about selecting $\lambda$ are given in Algorithm 1 in Section \ref{algorithm}.

\subsection{Auxiliary importance sampling}\label{PSML_auxiliary}
The first class of importance samplers with auxiliary parameter $\rho$ in \eqref{h_rho} is given by 
\begin{align}\label{auxiliary_importance}
q_\rho(\boldsymbol{X}^{m+1}|\boldsymbol{X}^m,&\boldsymbol{X}_{\text{obs}}(t_i))=\phi(\boldsymbol{X}^{m+1}; \boldsymbol{X}^m+\boldsymbol{\eta}_m\delta,\rho\Sigma_m\delta),
\end{align}
where $\boldsymbol{\eta}_m, \Sigma_m$ are defined in \eqref{MBB_eta} and $\eqref{MBB_sigma}$. Hence, $\rho\in[0,1]$ is the shrinkage coefficient, which will be estimated as an auxiliary parameter in the penalized log likelihood \eqref{pen_max_lambda}. The penalty term in \eqref{pen_max} allows estimation of the auxiliary parameter $\rho$, which is a feature of PSML, and leads to improved performance over the MBB and regularized sampler as will be illustrated in Section \ref{simulation}. Note that the MBB sampler is a special case of our auxiliary importance sampler \eqref{auxiliary_importance} when $\lambda=0$ and $\rho=1$.

We consider the regularized sampler \eqref{regularized_sampler} as another class of auxiliary importance samplers with auxiliary parameter $\rho$. The optimal choice of $\rho$ can be determined by maximizing the penalized log likelihood \eqref{pen_max}. As will be shown below in Section \ref{simulation}, this leads to improved performance of the regularized sampler as compared to fixing $\rho=0.1$ as in \citet{lindstrom2012regularized}.

One can also choose other families of auxiliary importance samplers, but the two classes considered above are a good starting point for illustration of the method. Other distributions, such as the Student's $t$ distribution, might also be a suitable choice.

\subsection{Algorithm for PSML}\label{algorithm}
The Algorithm for penalized simulated maximum likelihood estimation is given in Algorithm 1. We consider two stopping criteria for this algorithm, $\epsilon_0$ and $\delta_\epsilon$. First, $\epsilon_0$ monitors the estimated prediction error $\epsilon_\lambda$ in \eqref{Estimated_PE}. If the estimated prediction error is sufficiently small, $\epsilon_\lambda\leq\epsilon_0$, then there is no need to tune $\lambda$ and the algorithm stops. The criterion $\delta_\epsilon$ monitors the change in the estimated prediction error. If the improvement is small or there is no improvement at all, that is $\epsilon_\lambda-\epsilon_{\lambda_*}\leq\delta_\epsilon$, then the algorithm stops.

\RestyleAlgo{boxruled}
\begin{algorithm}[ht]
\caption{Algorithm for penalized simulated maximum likelihood estimation.}
\noindent\fcolorbox{white}{white}{\parbox{0.95\textwidth}{      
\begin{enumerate}[Step 1.]
\item Pick $\lambda_0>0$, $\epsilon_0>0$, $\delta_\lambda>0$, and $\delta_\epsilon>0$. Let $\lambda=\lambda_0$.
\item Find the maximizer $(\hat{\boldsymbol{\theta}},\hat{\rho})$ of the penalized log likelihood in \eqref{pen_max_lambda}. Compute the estimated prediction error $\epsilon_\lambda$ in \eqref{Estimated_PE}.
\item If $\epsilon_\lambda<\epsilon_0$ then stop, otherwise go to Step 4.
\item Let $\lambda_*=\lambda-\delta_\lambda$. Compute $\epsilon_{\lambda_*}$. 
\item If $\epsilon_\lambda-\epsilon_{\lambda_*}>\delta_\epsilon$ then update $\lambda=\lambda_*$ and go back to Step 3, otherwise go to Step 6.
\item If $\lambda<\lambda_0$ (i.e. $\lambda$ was updated from the initial $\lambda_0$) then stop, otherwise go to Step 7.
\item If $\epsilon_{\lambda}<\epsilon_0$ then stop, otherwise go to Step 8.
\item Let $\lambda_*=\lambda+\delta_\lambda$. Compute $\epsilon_{\lambda_*}$ as in Step 2. 
\item If $\epsilon_\lambda-\epsilon_{\lambda_*}>\delta_\epsilon$ then update $\lambda=\lambda_*$ and go back to Step 7, otherwise stop.
\end{enumerate}
}}
\end{algorithm}

For Step 1, we find that the initial value $\lambda_0\in(0.1, 0.5)$ works well for our models considered in Section \ref{simulation}. The values $\epsilon_0$ and $\delta_\epsilon$ are data dependent. The parameter $\delta_\lambda$ is the step size for exploring the space of $\lambda$ values. We use $\delta_\lambda=0.025$.

Note that, although this procedure looks computationally intensive, the algorithm converges quickly and is robust to the choice of $\lambda_0$ (as will be illustrated in the simulation studies in Section \ref{simulation}). Based on our simulation studies, we find that the first three steps, Steps 1 to 3, in the procedure are already sufficient to gain an improvement over the MBB sampler or the regularized sampler.

Note that Algorithm 1 can be extended to a parallel procedure by repeating Step 2 through a grid search with grid width $\delta_\lambda$ on the interval $\lambda\in (0, c)$, where $c$ is a constant. In our experience $\lambda\in(0,1)$ is reasonable. In this case, the parameter estimates $(\hat{\boldsymbol{\theta}},\hat{\rho})$ that correspond to the smallest $\epsilon_\lambda$ would be the output.

We use the parametric bootstrap \citep[Chapter~5]{efron1982jackknife} to obtain confidence intervals for the estimator, which proceeds as follows. First, based on the parameter estimates from the original dataset of interest, we can generate a large number of new \emph{datasets} by using the Euler-Maruyama scheme for the SDE model \eqref{general_SDE_euler}. For each new simulated dataset, we obtain estimates of parameters using the PSML method described in Algorithm 1. Then we compute the confidence interval from those estimates using the corresponding quantiles.

\section{Simulation studies}\label{simulation}
Here, we compare the performances of the MBB sampler, the regularized sampler with $\rho=0.1$, and our PSML with the modified MBB class \eqref{auxiliary_importance} and the regularized class \eqref{regularized_sampler} on simulated datasets for three different models. We refer to PSML with the modified MBB class \eqref{auxiliary_importance} as PSML-MBB and refer to PSML with the regularized class \eqref{regularized_sampler} as PSML-Reg. For all the optimization algorithms in this paper, we use an implementation of the Nelder-Mead algorithm for derivative-free optimization \citep{dfoptim2011} in R \citep{Rcitation} on an Intel Xeon W3565 3.2 GHz with CentOS 6 computer. The iterations for optimization of Step 2 of Algorithm 1 will stop when the absolute difference in function values between successive iterations is below $10^{-6}$, which is the default value in the R dfoptim package. We also use the default value for the maximum number of objective function evaluations allowed, which is 1500 for all three models below. The initial values for the parameters are chosen arbitrarily (we tried different initial values and obtained similar results). No parallel algorithm is involved in all the reported computation times. The time to compute the confidence intervals is not included. 

\subsection{Ornstein-Uhlenbeck process}
We first consider a univariate SDE, the Ornstein - Uhlenbeck process
\begin{equation}\label{OU_process}
dX=(\theta_1-\theta_2 X)dt+\theta_3 dW,
\end{equation}
with known initial condition $X(t_0)$, and the parameter $\boldsymbol{\theta}=(\theta_1,\theta_2,\theta_3)\in\mathbb{R}\times\mathbb{R}_{+}\times\mathbb{R}_{+}$. The parameter $\theta_2$ is the speed of reversion, $\theta_1/\theta_2$ is the long-run equilibrium value of the process, and $\theta_3$ is interpreted as the volatility. We generate 100 datasets, each including 100 observations, with initial condition $X(t_0)=1$, the time interval $t_i-t_{i-1}=1$, and parameter $\boldsymbol{\theta}_0=(0.0187, 0.2610, 0.0224)$ as reported in \citet{ait2002transition}. 

The transition density between two observations is given by
\begin{align*}
&X(t_{i+1})|X(t_i)\sim N\left(\frac{\theta_1}{\theta_2}+\left(X(t_i)-\frac{\theta_1}{\theta_2}\right)e^{-\theta_2\Delta}, \frac{\theta_3^2(1-e^{-2\theta_2\Delta})}{2\theta_2}\right),
\end{align*}
where $\Delta=t_{i+1}-t_i$ for $i=1,\dots,n-1$ \citep[Chapter~3]{iacus2009simulation}. Hence, the exact likelihood is known for this case and we can obtain the exact maximum likelihood estimator of the parameters $\boldsymbol{\theta}$. We compute the bias and the root mean square error
(RMSE) of the simulated maximum likelihood estimators $\widehat{\boldsymbol{\theta}}_r$ with respect to the exact maximum likelihood estimators $\widehat{\boldsymbol{\theta}}_\text{MLE}$, defined by $\frac{1}{100}\sum_{r=1}^{100}(\widehat{\boldsymbol{\theta}}_r-\widehat{\boldsymbol{\theta}}_\text{MLE})$ and $\sqrt{\frac{1}{100}\sum_{r=1}^{100}(\widehat{\boldsymbol{\theta}}_r-\widehat{\boldsymbol{\theta}}_\text{MLE})^2}$, respectively. For all the methods, we consider $M=8$ subintervals but with different levels of the number of simulated sample paths $J$. We set the $\epsilon_0=0.04$, $\lambda_0=0.25$, $\delta_\lambda=0.025$ and $\delta_\epsilon=0.001$. The initial values for optimization for $\theta_1, \theta_2$, and $\theta_3$ are $0.05, 0.5$ and $0.05$, respectively.

\begin{table}
\centering
\caption{The bias and RMSE of the simulated maximum likelihood estimates with respect to the exact maximum likelihood estimates for the Ornstein-Uhlenbeck process \eqref{OU_process}. All results are multiplied by $10^4.$ Both PSML-MBB and PSML-Reg have better performance than the MBB sampler and the regularized sampler in terms of reducing bias and RMSE, especially when the number of sample paths is small ($J=8$).}
\label{OU_result}
\begin{tabular}{l | l | r  r  r | r r  r }
& & \multicolumn{3}{c|}{$J=8$} & \multicolumn{3}{c}{$J=16$}\\
& Method & $\theta_1 $ & $\theta_2$ & $\theta_3$ & $\theta_1$ & $\theta_2$ & $\theta_3$\\ \hline
\multirow{4}{*}{Bias $(\times10^{-4})$} 
& MBB & 46 & 382  & 35 & 17 & 121 & 12 \\
& Regularized & 12 & 85 & 8 & 7 & 37 & 4 \\
& \textbf{PSML-MBB} & $-6$ & $-69$ & $-5$ & $-8$ & $-77$ & $-5$ \\
& PSML-Reg & $-2$ & $2$ & 3 & $-5$ & $-39$ & 5\\\hline 
\multirow{4}{*}{RMSE $(\times10^{-4})$} 
& MBB & 115 & 982  & 93 & 74 & 585 & 57 \\
& Regularized & 67 & 505 & 49 & 52 & 406 & 43\\
& \textbf{PSML-MBB} & 16 & 114 & 10 & 15 & 105 & 8\\
& PSML-Reg & 21 & 142 & 15 & 22 & 135 & 13  \\
\end{tabular}
\end{table}

Table \ref{OU_result} shows that both PSML-MBB and PSML-Reg have better performance than the MBB sampler and the regularized sampler in terms of reducing bias and RMSE, especially when the number of sample paths is small ($J=8$). We find that more accurate estimates can be achieved by introducing the penalty term in the PSML and selecting the optimal $\rho$ for the regularized class, which is in contrast to the fixed $\rho$ case for the regularized sampler as studied in \citet{lindstrom2012regularized}. 

\begin{figure}
\centerline{\includegraphics[width=0.9\textwidth]{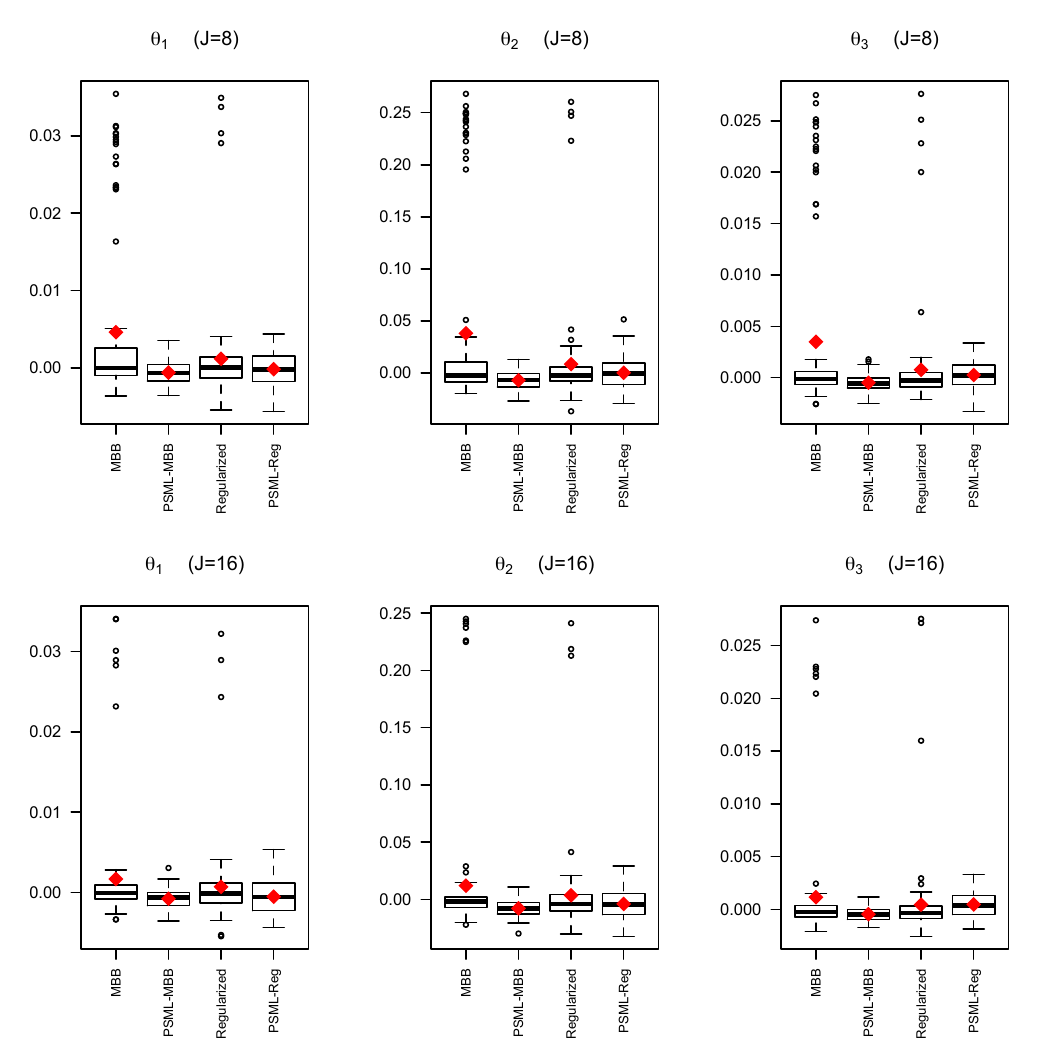}}
\caption{Boxplot of the bias of 100 estimates with $J=8$ and $J=16$ for the Ornstein-Uhlenbeck process \eqref{OU_process}. The red bold diamond points are the mean. Note that the PSML greatly reduces the Monte Carlo variability of the estimates.}
\label{OU_boxplot}
\end{figure}

Figure \ref{OU_boxplot} shows that some estimates are far away from the exact maximum likelihood estimates for the MBB sampler and the regularized sampler. This typically happens when the estimates based on the log likelihood approximated by the MBB sampler or the regularized sampler get stuck at the local maxima for optimization. This may be an indication of the poor approximation of the likelihood, since as the number of sample paths $J$ increases, fewer estimates have large bias.  For a small $J$ a poor choice of proposal distribution, e.g., the MBB and the regularized sampler, could result in a poor approximation to the likelihood because the approximated likelihood surface is generally more wiggly when $J$ is small. Thus when $J$ is small it is more common for these methods to mistakenly select a local maximum. The proposed PSML-MBB and PSML-Reg have better performance in this regard.

\begin{figure}
\centerline{\includegraphics[width=0.9\textwidth]{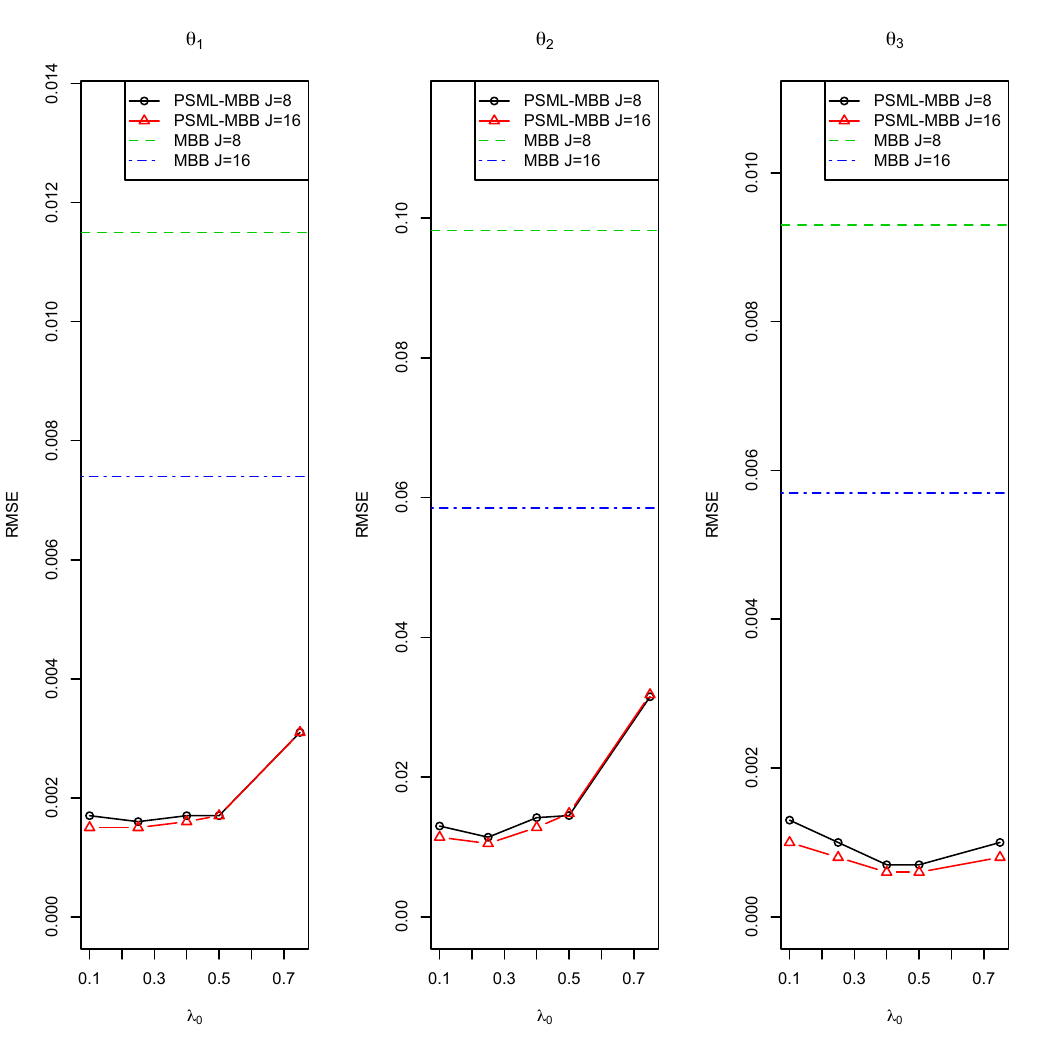}}
\caption{The RMSE of the MBB and PSML-MBB estimates with different $\lambda_0$ for the Ornstein-Uhlenbeck process \eqref{OU_process}.}
\label{OU_diff_lambda}
\end{figure}

Figure \ref{OU_diff_lambda} indicates that the performance of PSML is robust to the choice of $\lambda_0$. The improvements of the PSML-MBB with various $\lambda_0$'s over the MBB sampler are similar. This makes the algorithm easy to implement in practice. Clearly, when $J$ is small the difference between the PSML-MBB and the MBB is very large. As the number of sample paths $J$ increases, the difference between the MBB sampler with the PSML-MBB decreases. However, this is not always true for other SDE models. See the Lorenz model \eqref{Lorenz} in the next section for more details.

To obtain a similar level of accuracy as the PSML-MBB with $J=8$, the MBB sampler requires at least $J=96$. However, the PSML-MBB with $J=8$ requires much less time (around 1/5) than the MBB with $J=96$. For the computation time, the PSML with $J=8$ takes 90 -- 110 seconds and 180 -- 200 seconds for $J=16$ (for both PSML classes in Section \ref{PSML_auxiliary}). The MBB sampler or the regularized sampler takes 65 -- 75 seconds to implement for $J=8$, 120 -- 140 seconds for $J=16$, and 750 -- 950 seconds for $J=96$. The computational time grows approximating linearly in $J$ for both algorithms. 

For the PSML-MBB with $\lambda_0=0.25$, the mean of $\hat{\rho}$ equals 0.94 and the mean of $\hat{\lambda}$ equals 0.24. Note that though $\hat{\rho}$ is close to 1 for the PSML-MBB, the MBB is equivalent to the PSML-MBB only when $\rho=1$ \emph{and} $\lambda=0$. For the PSML-Reg with $\lambda_0=0.25$, the mean of $\hat{\rho}$ equals 0.33 and the mean of $\hat{\lambda}$ equals 0.23. We note that the performance of PSML-Reg presented in Tables \ref{OU_result} -- \ref{CWD_sim_result} is based on the estimated $\hat{\rho}$. (We fix $\rho=0.1$ for the regularized sampler as in \citet{lindstrom2012regularized}.) This indicates the regularized sampler can be improved when $\rho$ is estimated in \eqref{regularized_sampler}. We have observed similar $\hat{\rho}$ values for the other models in the simulation studies considered in this section.

\subsection{Stochastic Lorenz 63 model}
Next, we consider the stochastic version of the well-known chaotic Lorenz 63 model \citep{lorenz1963deterministic, bengtsson2003toward}, which is given by 
\begin{equation}\label{Lorenz}
d\begin{pmatrix}
X_1\\
X_2\\
X_3
\end{pmatrix}=\begin{pmatrix}
s(X_2-X_1)\\
rX_1-X_2-X_1X_3\\
X_1X_2-bX_3
\end{pmatrix}dt+\sigma d
\begin{pmatrix}
W_1\\
W_2\\
W_3
\end{pmatrix},
\end{equation}
where $W_1$, $W_2$, and $W_3$ are three independent Wiener processes. 

We again generate 100 datasets, each including 21 observations, with initial condition $(-10, -10, 30)$, time interval $t_i-t_{i-1}=0.05$, and commonly used parameter values $\boldsymbol{\theta}_0=(s_0=10,r_0=28,b_0=8/3,\sigma_0=2)$ \citep{bengtsson2003toward}. We assume all state variables, $(X_1, X_2, X_3)^T$, are observed at $t_i$ for $i=0, \dots, n$. In this case, the exact transition density is no longer available. We can only compute the bias and the RMSE of the simulated maximum likelihood estimators $\widehat{\boldsymbol{\theta}}_r$ with respect to the true parameters $\boldsymbol{\theta}_0$, defined by $\frac{1}{100}\sum_{r=1}^{100}\allowbreak(\widehat{\boldsymbol{\theta}}_r-\boldsymbol{\theta}_0)$ and $\sqrt{\frac{1}{100}\sum_{r=1}^{100}(\widehat{\boldsymbol{\theta}}_r-\boldsymbol{\theta}_0)^2}$, respectively. Estimates are obtained using different sample paths $J$ and $M=10$ subintervals. We set $\lambda_0=0.5$, $\epsilon_0=3.5$, $\delta_\lambda=0.025$ and $\delta_\epsilon=0.1$. The initial values for optimization are $(15,30,5,1)$.

\begin{table}
\centering
\caption{The bias and RMSE of the simulated maximum likelihood estimates with respect to the true parameters for the stochastic Lorenz 63 model \eqref{Lorenz}. The improvement of the PSML-Reg over the MBB and the regularized sampler with different $J$'s is evident.}
\label{Lorenz_result}
\begin{tabular}{ l | l | r r r r }
&Method & $s$ & $r$ & $b$ & $\sigma$  \\ \hline 
\multirow{5}{*}{Bias} 
& MBB ($J=128$) & 1.86 & $-$15.70 & 8.92 & 9.59 \\ 
& Regularized ($J=32$) & 2.59 & $-$0.25 & $-$0.03 & 1.54 \\ 
& Regularized ($J=48$) & 3.79 & $-$0.07 & $-$0.06 & 1.25 \\ 
& Regularized ($J=64$) & 3.86 & $-$0.07 & 0.00 & 1.17 \\ 
& Regularized ($J=128$) & 3.89 & $-$0.02 & $-$0.03 & 0.99 \\ 
& \textbf{PSML-Reg} ($J=32$) &$-$1.75 & 0.00 & 0.01 & 0.38\\\hline
\multirow{5}{*}{RMSE} 
& MBB ($J=128$) & 13.31 & 20.70 & 15.52 & 15.36 \\ 
& Regularized ($J=32$) & 18.56 & 1.23 & 0.39 & 3.70 \\ 
& Regularized ($J=48$)  & 21.82 & 0.81 & 0.45 & 3.26 \\
& Regularized ($J=64$) & 20.57& 0.59 & 0.28 & 2.76 \\  
& Regularized ($J=128$) & 21.79 & 0.51 & 0.30 & 2.54 \\ 
& \textbf{PSML-Reg} ($J=32$) & 3.54 & 0.31 & 0.10 & 0.54\\
\end{tabular}
\end{table}

\begin{figure}
 \centerline{\includegraphics[width=0.6\textwidth]{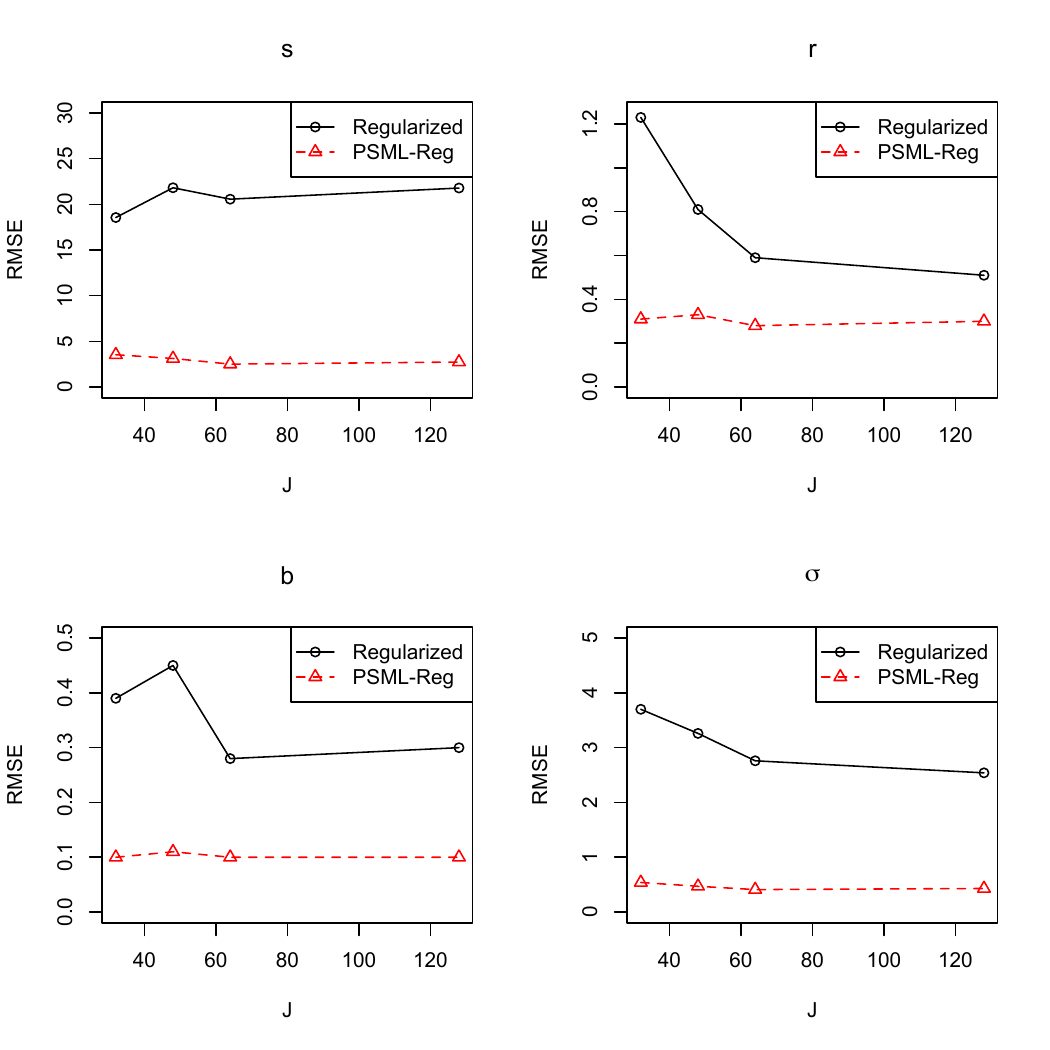}}
\caption{The RMSE of the regularized and the PSML-Reg estimates with different $J$ for the stochastic Lorenz 63 model \eqref{Lorenz}.}
\label{Lorenz_diff_J}
\end{figure}

As shown in Table \ref{Lorenz_result} the MBB sampler performs poorly for the stochastic Lorenz 63 model \eqref{Lorenz}, especially for the parameters $r,b$, and $\sigma$ as \citet{lindstrom2012regularized} has observed. This is because the dynamics of the Lorenz model is dominated by the drift term. The MBB sampler ignores the dynamics of the model and generates paths far from the actual realization. Moreover, Figure \ref{Lorenz_diff_J} shows that there is no significant increasing trend in the accuracy of the regularized sampler as the number of sample paths $J$ increases, especially for parameter $s$. A large $J$ but a fixed $\rho$, which controls the weight between the Pedersen and the MBB sampler, still cannot assure the regularized sampler generates paths close to the actual trajectories. However, the improvement of the PSML-Reg over the MBB and the regularized sampler is evident. An estimated $\hat{\rho}$ based on the data in the PSML-Reg plays an important role in generating efficient proposal trajectories. 

In terms of computational time, both the MBB and the regularized samples need 1800 -- 2100 seconds for $J=32$, 2700 -- 3000 seconds for $J=48$, 3500 -- 4000 seconds for $J=64$, and 7000 -- 8000 seconds for $J=128$. The PSML-MBB with $J=32$ requires 2000 -- 2300 seconds.

\subsection{CWD direct transmission model}
The specific model and background are described in Section 6. Again, we generate 100 datasets, each including 21 annual observations from two distinct CWD epidemics similar to the real dataset in Section 6, by using the CWD direct transmission model \eqref{CWD_direct} with parameter $(\beta_0=0.03,\mu_0=0.20)$. The initial condition $\boldsymbol{X}(t_0)=(S(t_0),\allowbreak I(t_0),C(t_0))^T$ is set to be the same as the real dataset. The step size of the Euler-Maruyama scheme is 1/12 of the time interval between each pair of observations, which is one month in this case. We set $\lambda_0=0.5$, $\epsilon_0=5$, $\delta_\lambda=0.025$ and $\delta_\epsilon=0.5$. The initial values for optimization for $\beta$ and $\mu$ are 0.05 and 0.5, respectively. Parameter estimates are obtained using $J=72$ sample paths for the MBB sampler and the regularized sampler, which require 2400 -- 2700 seconds, and $J=48$ for the PSML-MBB and the PSML-Reg, which require 2000 -- 2300 seconds. The exact transition density is not available for this case. The bias and RMSE of the simulated maximum likelihood estimates with respect to the true parameters are shown in Table \ref{CWD_sim_result}, which indicate similar improvements of the PSML-MBB and the PSML-Reg over the MBB sampler and the regularized sampler. For this simulation the states $S$ and $I$ are unobserved and the time between observations is long (yearly). The fact that the PSML does well in this context is promising for this and other applications in ecology. Since the best results are obtained by the PSML-MBB with $J=48$ sample paths, we use the same setting in the real data example in Section \ref{CWD}.

\begin{table}
\centering
\caption{The bias and RMSE of the simulated maximum likelihood estimates with respect to the true parameters for CWD direct transmission model \eqref{CWD_direct}. Both PSML-MBB and PSML-Reg have better performance than the MBB sampler and the regularized sampler.}
\label{CWD_sim_result}
\begin{tabular}{ l | l | r  r }
&Method & $\beta$ & $\mu$  \\ \hline 
\multirow{4}{*}{Bias} 
& MBB ($J=72$) & 0.02 & 0.07\\
& Regularized ($J=72$) & 0.01 & 0.07 \\ 
& \textbf{PSML-MBB} ($J=48$) & 0.01 & 0.02   \\
& PSML-Reg ($J=48$) & 0.01 & 0.04\\\hline
\multirow{4}{*}{RMSE} 
& MBB ($J=72$) &0.07 & 0.11 \\
& Regularized ($J=72$)&0.04 & 0.12\\ 
& \textbf{PSML-MBB} ($J=48$) & 0.02 & 0.05\\
& PSML-Reg ($J=48$) & 0.02 & 0.06
\end{tabular}
\end{table}

\section{Chronic wasting disease example}\label{CWD}
Deer populations and ecosystems can be severely disrupted by the contagious prion disease, known as chronic wasting disease (CWD) \citep{miller2006dynamics}. In order to reduce the potential damages caused by CWD, it is important to understand the transmission mechanisms of CWD. Several deterministic epidemic models were proposed by \citet{miller2006dynamics} in order to portray the transmission of CWD. Here, based on one of those deterministic models, we firstly derive a CWD SDE model using the technique described in \citet[Chapter~8]{allen2003introduction}. Then, we implement the proposed PSML method to the dataset studied in \citet{miller2006dynamics}. Their dataset consists of annual observations of cumulative mortality from two distinct CWD epidemics (Figure \ref{CWD_direct_fig} upper display) in captive mule deer held at the Colorado Division of Wildlife Foothills Wildlife Research Facility in Fort Collins, Colorado. The first epidemic occurred from 1974 to 1985 and the second epidemic occurred in a new deer herd from 1992 to 2001. The dataset also includes the annual number of new deer added to the herd and the per capita losses due to natural deaths and removals. We note that the dataset contains no measurement or observation error since it was recorded in a captive laboratory facility. We assume the direct transmission coefficient $\beta$ and the per capita CWD mortality rate $\mu$ do not change between two epidemics as such parameters are innate characteristics of the associated disease. Hence we can combine two epidemics as a single dataset for estimating the parameters.

\subsection{CWD direct transmission model}
CWD may be transmitted to susceptible animals directly from infected animals \citep{miller2006dynamics}. We portray this direct transmission using an SDE model. Let $\boldsymbol{X}(t)=(S(t),I(t),C(t))^T$, where $S$ is the number of susceptible animals, $I$ is the number of infected animals, $C$ is the total number of accumulate deaths from CWD over time. We assume the initial condition $\boldsymbol{X}(t_0)=(S(t_0),I(t_0),C(t_0))^T$ is known. Also, our basic assumption is that only $C$ can be observed at $t_i$, for $i=1,\ldots,n$, and the other two state variables, $S$ and $I$, are unobserved.  The unknown parameters to be estimated in the epidemic model are denoted by $\boldsymbol{\theta}=(\beta,\mu)$, where $\beta$ is the direct transmission coefficient, $\mu$ is the per capita CWD mortality rate. Then the direct transmission SDE model is given by
\begin{equation}\label{CWD_direct}
d\begin{pmatrix}
S\\
I\\
C
\end{pmatrix}=\begin{pmatrix}
a-S(\beta I+m)\\
\beta SI-I(\mu+m)\\
\mu I
\end{pmatrix}dt+\boldsymbol{B}d\boldsymbol{W}
\end{equation}
where $a$ is the known number of susceptible animals annually added to the population via births or importation, $m$ is the known per capita natural mortality rate, $\boldsymbol{W}=(W_1,W_2,W_3)^T$ is a 3-dimensional standard Wiener process, and $\boldsymbol{B}=\sqrt{\Sigma}$ is the positive definite square root of the covariance matrix with
\begin{equation}\label{Sigma_direct}
\Sigma=\begin{bmatrix}
a+S(\beta I+m) & -\beta SI & 0\\
-\beta SI & \beta SI+I(\mu+m) & -\mu I\\
0 & -\mu I & \mu I
\end{bmatrix}.
\end{equation}

Although the SDE model \eqref{CWD_direct} relaxes the assumption of discrete states and non-negative nature of $S$, $I$, and $dC$, similar SDE models have been used to approximate the transmissions of epidemics in several recent articles \citep{ionides2006inference, bhadra2011malaria, golightly2011bayesian}. We also monitor the frequency of estimates in $S$, $I$, and $dC$; they were rare to the point of negligibility in our analysis.

Here, we briefly explain how the above SDE model is derived. See \citet[Chapter~8]{allen2003introduction} for more details. Let $\boldsymbol{X}_\delta=\boldsymbol{X}(t+\delta)-\boldsymbol{X}(t)$ be the increment during the time interval $\delta$. If $\delta$ is sufficiently small, we can assume at most one animal is infected or died during the time interval $\delta$. The probability of an event that more than one infection or death has occurred during time $\delta$ is of order $\delta^2$, which can be neglected. Then we can approximate the mean of $\boldsymbol{X}_\delta$ for $\delta$ sufficiently small to order $\delta$ by
\begin{equation}\label{direct_expectation}
E[\boldsymbol{X}_\delta]\approx f\delta=
\begin{pmatrix}
a-S(\beta I+m)\\
\beta SI-I(\mu+m)\\
\mu I
\end{pmatrix}\delta.
\end{equation}
Furthermore, we can also approximate the covariance of $\boldsymbol{X}_\delta$ for $\delta$ sufficiently small by 
\begin{align}\label{direct_variance}
V[\boldsymbol{X}_\delta] &= E[(\boldsymbol{X}_\delta)(\boldsymbol{X}_\delta)^T]-E(\boldsymbol{X}_\delta)E(\boldsymbol{X}_\delta)^T\approx E[(\boldsymbol{X}_\delta)(\boldsymbol{X}_\delta)^T]=\Sigma\delta.
\end{align}
The matrix $\Sigma$ in \eqref{Sigma_direct} is positive definite and hence has a positive definite square root $\boldsymbol{B}=\sqrt{\Sigma}$. It can be shown that \eqref{direct_expectation} and \eqref{direct_variance} are quantities of order $\delta$. We also assume $\boldsymbol{X}_\delta$ follows normal distribution with mean vector $f\delta$ and covariance matrix $\boldsymbol{B}^2\delta=\Sigma\delta$. Thus,
\begin{equation}\label{direct_explanation}
\boldsymbol{X}(t+\delta)\approx\boldsymbol{X}(t)+f\delta+\boldsymbol{B}\sqrt{\delta}\boldsymbol{\eta},
\end{equation}
where $\boldsymbol{\eta}\sim N(0, \boldsymbol{\mathcal{I}}_{3\times 3})$ and $\boldsymbol{\mathcal{I}}$ is the identity matrix. This is exactly one iteration of the Euler-Maruyama scheme for a system of SDEs \eqref{CWD_direct}. As a result, the dynamical system \eqref{direct_explanation} converges in the mean square sense to the system of SDEs \eqref{CWD_direct} as $\delta\rightarrow 0$.

\subsection{Results}
The simulated maximum likelihood estimates based on the PSML-MBB with $J=48$, and $M=12$ are $\widehat{\boldsymbol{\theta}}_{\text{PSML-MBB}}\allowbreak=(\hat{\beta},\hat{\mu})=(0.03, 0.21)$ with 95\% confidence intervals $[0.027, 0.120]$ and $[0.143, 0.388]$, respectively, and $\hat{\rho}=0.86$. Although Durham and Gallant (2002) did not provide confidence intervals based on the MBB approach or a method to compute them, we use the parametric bootstrap as described in Section \ref{algorithm} to obtain them. The estimates based on the MBB approach with $J=48$ are $\widehat{\boldsymbol{\theta}}_{\text{MBB}}=(0.03,0.27)$ with 95\% confidence intervals $[0.027,0.186]$ and $[0.148, 0.599]$, which are much wider than those from the PSML-MBB. 

\begin{figure}
 \centerline{\includegraphics[width=0.55\textwidth]{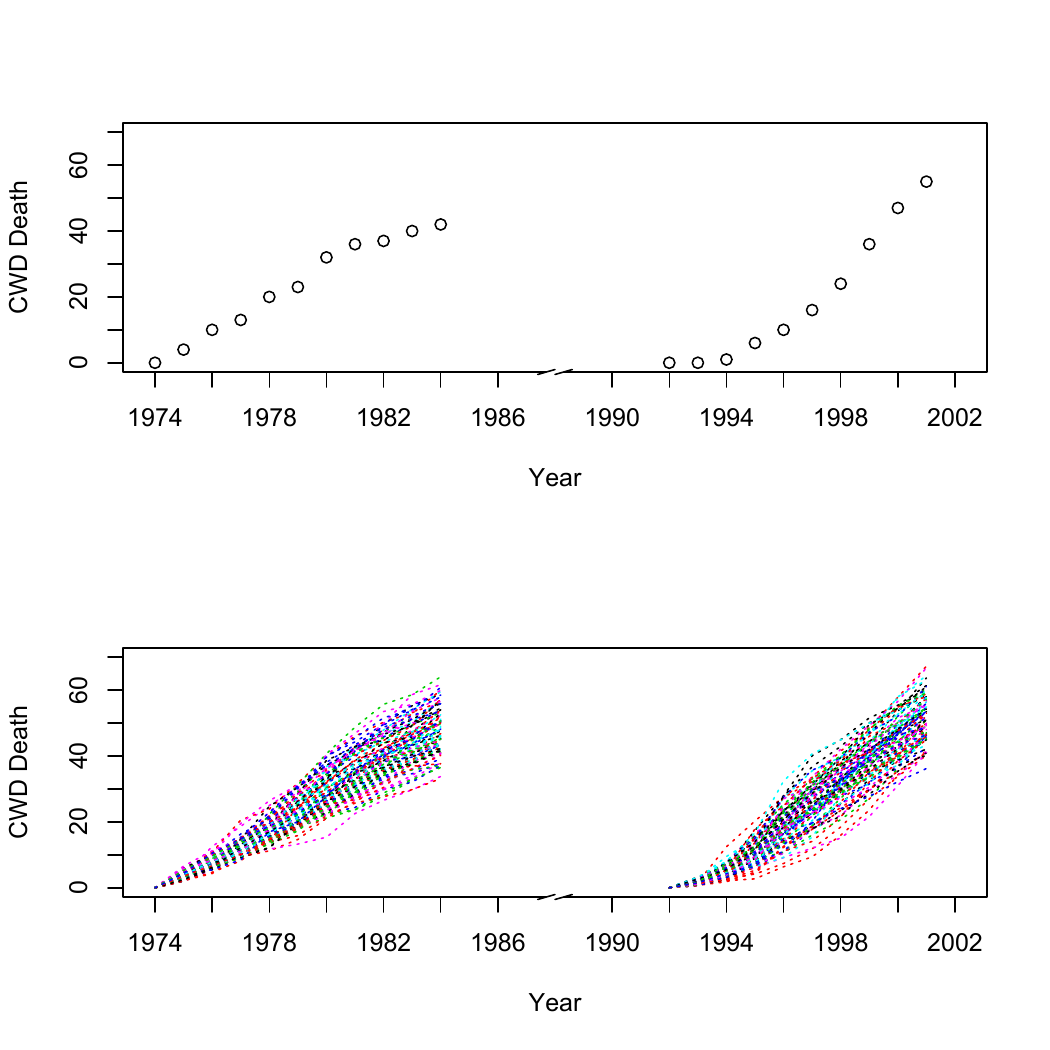}}
\caption{Upper display: observed cumulative number of deaths for CWD. Lower display: the 100 simulated trajectories of the cumulative number of deaths for CWD are obtained by using CWD direct transmission model \eqref{CWD_direct} with estimated parameters from the PSML-MBB. The circled points are the observed CWD data.}
\label{CWD_direct_fig}
\end{figure}

To measure the goodness of fit, 100 simulated trajectories of cumulative number of deaths for CWD using $\widehat{\boldsymbol{\theta}}_{\text{PSML-MBB}}$ are shown in Figure \ref{CWD_direct_fig}. For such a small sample size the estimated parameters from the PSML-MBB and the CWD direct transmission model capture the pattern of the CWD death data over time. The fit for the second epidemic is not as good as the first epidemic because we are estimating parameters ($\mu$ and $\beta$, which remain unchanged between epidemics) from a theoretical SDE model \eqref{CWD_direct}, not estimating a least squares fit to the observed data. The theoretical model does a remarkably good job at following the observed data. A non-parametric model would likely provide a close fit to the data in Figure \ref{CWD_direct_fig} but would not provide the scientifically relevant interpretation sought by biologists. \citet{miller2006dynamics} proposed a more complex deterministic model, which we could also extend to a corresponding stochastic model, however the model quickly becomes over-parameterized due to the limited sample size and complexity of the model. Therefore, we only consider the direct transmission model.

The basic reproductive number $R_0$, which is the average number of secondary cases generated by one infected individual over the course of its infectious period when the entire population is susceptible, is important in biology and epidemiology \citep{anderson1992infectious}. Usually people consider the situation in which the majority of a closed population is susceptible, that is $S(t_0)/N\approx1$. For deterministic models, if $R_0>1$ then the infection will be spread in a population, and if $R_0\leq1$, the infection will die out monotonically. For stochastic models, the probability that there is no epidemic equals 1 if $R_0\leq1$ and $(\frac{1}{R_0})^{I(t_0)}$ if $R_0>1$ \citep{allen2000comparison}, where $I(t_0)$ is the initial number of infected animals. The traditional interpretation of $R_0$ is not available here because the population is not closed; $a$ in \eqref{CWD_direct} is the known number of susceptible annually added to the population. However, we want to point out that our method can be used to estimate $R_0$ for cases when the population is closed and the other assumptions of $R_0$ hold. For example, assuming a natural mortality rate of $m=0.15$ \citep{miller2006dynamics}, the corresponding estimate for the basic reproductive number $R_0$ equals $\widehat{\beta} N_0/(\widehat{\mu}+m)\approx0.16N_0$ with 95\% confidence interval $[0.06N_0, 0.42N_0]$, where $N_0$ is the initial population or susceptible size. Hence, we would expect that CWD will spread if a few infected animals, like one or two, are introduced to a closed susceptible population with size at least $1/0.06\approx 17$ animals.

\section{Conclusion and discussion}
The dynamics of many ecological problems can be well described by a multivariate stochastic differential equation system. However, the transition densities of discrete-time observations are unknown for most interesting models. We propose the penalized simulated maximum likelihood approach, which provides a balanced approach to achieve accurate parameter estimates with efficient computation times for these complex stochastic models. The key idea is the introduction of a penalty term to select a better importance sampler in order to reduce the number of simulated sample paths. We compare the new method to the MBB sampler and the regularized sampler for three different models in simulation studies and also show an application for a real dataset. From those results, we conclude that the penalized simulated maximum likelihood approach is an improvement over the MBB sampler and the regularized sampler while keeping the computational cost low. 

Note that it is possible to extend the penalized simulated maximum likelihood approach to allow for measurement errors in our observed data. The main challenge is still constructing effective and efficient importance samplers to approximate the transition probability density. The detailed statistical procedures are left as further work. Alternative approaches, such as methods that do not require evaluation of the likelihood function, have been proposed in both frequentist \citep{breto2009time} and Bayesian analysis \citep{andrieu2010particle,sun2014ABC}.

Markov jump processes offer an alternative approach to using SDE models \citep{toni2009approximate, drovandi2011estimation}. A Markov jump model particularly takes into account the discreteness of the data. However, a Markov jump model may be too simple. For example, the SDE models considered here allow modeling of the covariance between state variables. In contrast, a Markov jump model cannot capture such a dependence structure among the state variables.

\citet{stramer2007asymptotics} concluded the optimal choice for the number of Monte Carlo simulations $J$ in \eqref{importance_sampler_MC} is of the order $O(M^2)$ for the MBB approach, where $M$ is the number of subintervals between two observations. One can choose a number smaller than this as a starting point for the proposed penalized simulated maximum likelihood method in practice. More formal guidance is under investigation. Moreover, a formal study about the tuning parameter $\lambda$ needs further development.

We find it is quite challenging to derive the theoretical properties of the maximum likelihood estimator based on either simulated likelihood (e.g., Pedersen and MBB) or penalized simulated likelihood (e.g., PSML). Pedersen (1995) and Geweke (1989) showed that the importance sampling estimator \eqref{importance_sampler_MC} converges to the transition density $p(\boldsymbol{X}(t_i)|\boldsymbol{X}(t_{i-1}))$. However, the properties of the MLE based on \eqref{importance_sampler_MC} (e.g., the estimator based on MBB or PSML) have not been established. The theoretical work, such as the convergence and the asymptotic distribution of the estimators, will be considered as future work. 

Note that uncertainty in $\hat{\rho}$ is not accounted for in the bootstrap confidence intervals for the SDE parameters.  This parameter is a nuisance parameter and is not used for simulated new datasets in the bootstrap algorithm.  Methods to account for the effect of estimating $\rho$ on bootstrap intervals for the process model parameters are a topic of future research.  

\section*{Acknowledgements}
This material is based upon work supported by the National Science Foundation under Grant No. EF-0914489 (Sun and Hoeting). The research work of Chihoon Lee is supported in part by the Army Research Office (W911NF-14-1-0216) and the National Security Agency (H98230-12-1-0250). This research also utilized the CSU ISTeC Cray HPS System, which is supported by NSF Grant CSN-0923386. We are grateful to N. Thompson Hobbs for introducing us to this problem, and Michael W. Miller and the Colorado Division of Parks and Wildlife for sharing the data. We would like to thank John Geweke and Garland Durham for stimulating conversations. We also appreciate the Associate Editor and the reviewers for their insightful suggestions that have greatly enhanced the manuscript.

\appendix
\section{Sequential Monte Carlo for stochastic differential equations with partially observed discrete data.}\label{append1}
A sequential Monte Carlo or particle filter scheme can be described as follows. Repeat the following steps for $i=1,\dots,n$,
\begin{enumerate}[(a).]
\item Sample $\boldsymbol{X}^{*(j)}_{-\text{obs}}(t_i)\sim q(\boldsymbol{X}_{-\text{obs}}(t_i)|\boldsymbol{X}^{(j)}(t_{i-1}),\boldsymbol{X}_{\text{obs}}(t_i))$ for $j=1,\dots,J$, where $\boldsymbol{X}^{(j)}(t_0)\equiv\boldsymbol{X}(t_0)$ and $q$ is an importance sampler.
\item Compute the weights $$\omega_i^{(j)}=\frac{p^{(M)}(\boldsymbol{X}^{*(j)}(t_i)|\boldsymbol{X}^{(j)}(t_{i-1}))}{q(\boldsymbol{X}_{-\text{obs}}^{*(j)}(t_i)|\boldsymbol{X}^{(j)}(t_{i-1}),\boldsymbol{X}_{\text{obs}}(t_i))},$$
and $W_i^{(j)}\propto\omega_i^{(j)}$, where $\boldsymbol{X}^{*(j)}(t_i)\equiv\{\boldsymbol{X}^{*(j)}_{-\text{obs}}(t_i),\boldsymbol{X}_{\text{obs}}(t_i)\}$ and $p^{(M)}$ in the numerator is defined in \eqref{importance_sampler}.
\item Resample $J$ times with replacement from $\{\boldsymbol{X}_{-\text{obs}}^{*(1)}(t_i),\dots,\boldsymbol{X}_{-\text{obs}}^{*(J)}(t_i)\}$ with probabilities given by $\{W_i^{(1)},\dots,W_i^{(J)}\}$ to obtain $J$ equally-weighted particles $\{\boldsymbol{X}_{-\text{obs}}^{(1)}(t_i),\dots,\boldsymbol{X}_{-\text{obs}}^{(J)}(t_i)\}$.
\end{enumerate}

\section*{References}
\bibliographystyle{elsarticle-harv} 
\bibliography{mybib}

\end{document}